\documentclass[conference]{IEEEtran}
\usepackage[T1]{fontenc}
\IEEEoverridecommandlockouts
\usepackage{cite}
\usepackage{amsmath,amssymb,amsfonts}
\usepackage{algorithmic}
\usepackage{graphicx}
\usepackage{textcomp}
\usepackage{xcolor}
\usepackage{float}
\usepackage{subfigure}
\usepackage{authblk}
\usepackage{tabularx}
\usepackage{balance}
\def\BibTeX{{\rm B\kern-.05em{\sc i\kern-.025em b}\kern-.08em
    T\kern-.1667em\lower.7ex\hbox{E}\kern-.125emX}}

\makeatletter
\newcommand{\linebreakand}{%
  \end{@IEEEauthorhalign}
  \hfill\mbox{}\par
  \mbox{}\hfill\begin{@IEEEauthorhalign}
}
\makeatother

\begin{document}

\title{Cross-platform graphics subsystem\\for an ARINC~653-compatible\\real-time operating system}

\author{Maksim Raenchuk\textsuperscript{1, 2}\\raenchuk@ispras.ru\\ORCID: 0000-0002-9901-0297 \and Vitaly Cheptsov\textsuperscript{1}\\cheptsov@ispras.ru\\ORCID: 0000-0003-3931-101X \and Alexey Khoroshilov\textsuperscript{1, 2}\\khoroshilov@ispras.ru\\ORCID: 0000-0002-6512-4632 \linebreakand \\ \textsuperscript{1}Ivannikov Institute for System Programming\\Moscow, Russia \and \\ \textsuperscript{2}Lomonosov Moscow State University\\Moscow, Russia}

\maketitle

\begin{abstract}
In the development of modern cockpits, there is a trend towards the use of large displays that combine information about air navigation and the status of aircraft equipment. Flight and equipment performance information generated by multiple flight control systems should be graphically displayed in an easy-to-read form on widescreen multifunction displays. It is usually generated by independent systems whose output must not interfere with each other in accordance with the requirements of the ARINC~653 standard. This paper presents a solution to the problem of displaying ARINC~653 applications, which further improves security and portability, when running multiple applications on a single screen of one physical device.
\end{abstract}

\begin{IEEEkeywords}
graphics subsystem, real-time operating system, widescreen multifunction displays
\end{IEEEkeywords}

\section{Introduction}
Different vendors may have varying levels of trust and software assurance. They should not be allowed to influence each other, even on the same platform. The lack of transparency of individual vendors in the area of secure development contributes to this, and forces the operating system to impose additional restrictions to achieve better stability. Also, certain industries may have to develop their software from scratch to adhere to specialised certification methods.

Currently, the main approach to the design and development of airborne systems of civil aircraft is the integrated modular avionics (IMA) approach \cite{b1}, which is used, among other things, in the Airbus A320, Boeing 787 Dreamliner, Sukhoi Superjet 100, and Irkut MC-21.

Integrated modular avionics is understood as the concept of building an onboard complex based on an open network architecture and a unified computing platform. According to this approach, specialised controllers are replaced by general-purpose processor modules, on which the independent operation of various aviation systems is ensured. A prerequisite for this capability is the distribution of time and resources between applications. It is precisely this mode of operation that a hard real-time operating system (RTOS) offers, which is an integral part and the most important component of the onboard system. JetOS is an airborne RTOS, developed to meet the ARINC~653 (Avionics Application Standard Software Interface) standard \cite{b2}, which serves exactly this purpose.

\begin{figure}[H]
    \centering
    \includegraphics[width=0.4\textwidth]{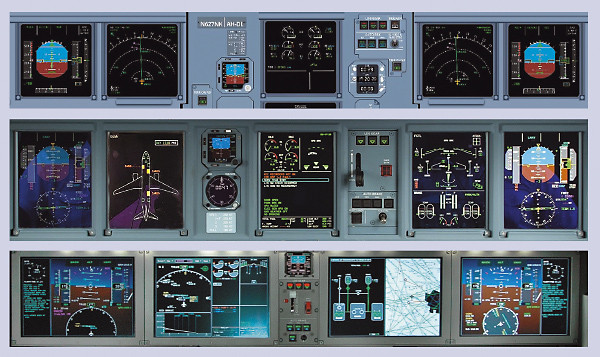}
    \caption{From top to bottom: Airbus A320, Sukhoi Superjet 100, and Irkut MC-21 cockpit panels \cite{b3}.}
\end{figure}

An important part of the onboard equipment of an aircraft is a system responsible for displaying information about the state of all aircraft systems in a form convenient for pilots to perceive.

The use of this concept and modern equipment (for example, computer displays instead of mechanical devices as one can see in the fig. 1) can significantly reduce the number of cables and devices onboard and, thus, simplify the structure, reduce the risk of breakage, and even reduce the take-off weight of the liner in some cases. Thus, there is a need to develop software for displays in the cockpit, which should provide a reliable interactive information source, using the resources of the low power processor, which is installed onboard.

\section{Background}

\subsection{Multi-window visualisation}

To visualise various sources of information in modern operating systems, the concept of a multi-window interface is used, when the contents of each application are displayed in a separate window. The approach taken in mission-critical systems is to fix application windows on the display. This is done for safety reasons, so that pilots, for example, do not accidentally move a window on the display in an overloaded aircraft.

To get a reasonable understanding of a multi-window representation in avionics, one can think of tiling window managers. These managers ``tile'' the windows so that none are overlapping. They usually make very extensive use of key bindings and have less (or no) reliance on the mouse, which is great for embedded aircraft systems. Tiling window managers may be manual, offer predefined layouts, or do both.

In avionics, there is also a concept of widgets that are standardised and allow to program the aircraft interface brick by brick. The ARINC~661 standard normalises the design of interactive cockpit display systems (CDS) and the way the CDS communicates with user applications (UA), including flight management, flight control and flight warning systems. It uses predefined and standardised graphical widgets, some changeable through pilot interaction via trackball, keyboard, tactile screens, by standardising the communication protocol at runtime between a UA and the CDS. ARINC~661 ensures that the full CDS interactively behaves with the avionics systems in the same manner, regardless of UA developer and CDS supplier.

Following this model, the graphical stack used in airborne systems is comprised of a graphical environment in each user application, server-side APIs for widget creation, which the applications interact with, widget rendering code, and a compositor, which job is to display all the rendered output of all applications on the target display.

\subsection{Assembling the graphical stack}

There are many different types of vendors that all have slightly different roles in supplying goods and services:
\begin{itemize}
    \item GPU vendors;
    \item GPU driver vendors:
    \begin{itemize}
        \item software rendering;
        \item hardware rendering (GPU vendors themselves);
    \end{itemize}
    \item operating system vendors;
    \item graphics library vendors for UA development.
\end{itemize}

When using hardware rendering the drawing operations (lines, circles, texts, bitmaps, transformations) are executed by the GPU. In contrast, when using software rendering the drawing operations are executed by the CPU, which is much slower. A CPU performs different calculations to process tasks while a GPU has the ability to focus all computing abilities on a specific task. A CPU is comprised of multiple cores (up to 24) that are used for sequential serial processing. A GPU utilises thousands of smaller and more efficient cores to handle multiple tasks simultaneously.

There are two specifications for safety-critical industries that define a platform-independent programming interface for writing 2D and 3D computer graphics applications: OpenGL SC \cite{b4} and Vulkan SC \cite{b5}.

The safety critical profile for OpenGL is defined to meet the unique requirements of the safety-critical applications such as avionics and automotive instrumentation displays. OpenGL SC 1.0 removes functionality from OpenGL ES 1.0 to minimise implementation and safety certification costs. It also adds functionality, such as display lists, that are required to support legacy and auto-generated display applications in safety critical markets.

Vulkan SC is a next-gen safety critical graphics API introduced by the Khronos Group. Vulkan is significantly more streamlined than OpenGL. It provides enhanced functionality with increased performance and flexibility, increased control of device scheduling, synchronisation, and resource management. Vulkan not only reduces CPU workload, but also provides multi-threading capabilities, and moves management of some functions to the application layer allowing more control with less overhead to provide performance gains. Vulkan is suited for a wide variety of applications currently supported by OpenGL.

Operating system vendors have to integrate GPUs, GPU drivers, and graphics libraries to enable UA development. The need for certification requires that the software development procedure meets standards and also requires full access to the source code of OpenGL implementation. For these reasons, many operating system vendors provide in-house graphic rendering solutions. This is true for Elbit Systems, Green Hills Software, or Mentor Graphics.

With the growth of graphics API standardisation, the implementations require more and more efforts on all sides, and companies providing multi-target certified drivers also start to appear. For example, Core Avionics and Industrial (CoreAVI) produces DO-178C and ISO 26262 certified drivers and libraries for the following list of GPUs \cite{b6}:
\begin{itemize}
    \item Radeon\textsuperscript{\texttrademark} E4690;
    \item Radeon\textsuperscript{\texttrademark} E8860;
    \item Radeon\textsuperscript{\texttrademark} E9171;
    \item Mali-G78AE.
\end{itemize}

However, the amount of supported hardware by such drivers is still rather limited, and the integration costs make an in-house solution cheaper or more efficient. It should also be taken into account that the certification of such drivers is nearly impossible without the participation of the graphics processor developers \cite{b7}.

\section{JetOS graphics support}

ARINC~653 is a standard that regulates the temporal and spatial division of resources in an airborne RTOS in accordance with the IMA principles to ensure that one application is protected from errors in another. So, when the processor switches from the execution of one application to the execution of another, the memory of the other partition becomes unavailable and the contents of registers and cache memory are necessarily cleared. Because of that, JetOS, as an ARINC~653-compatible operating system, is a particularly good choice for the use as an airborne graphics rendering platform.

To enable support of visual applications in JetOS the Keldysh Institute of Applied Mathematics (KIAM) develops a complete graphical/rendering subsystem both in software and in hardware, depending on the target certification level \cite{b8}.

\begin{figure}[H]
    \centering
    \includegraphics[width=0.4\textwidth]{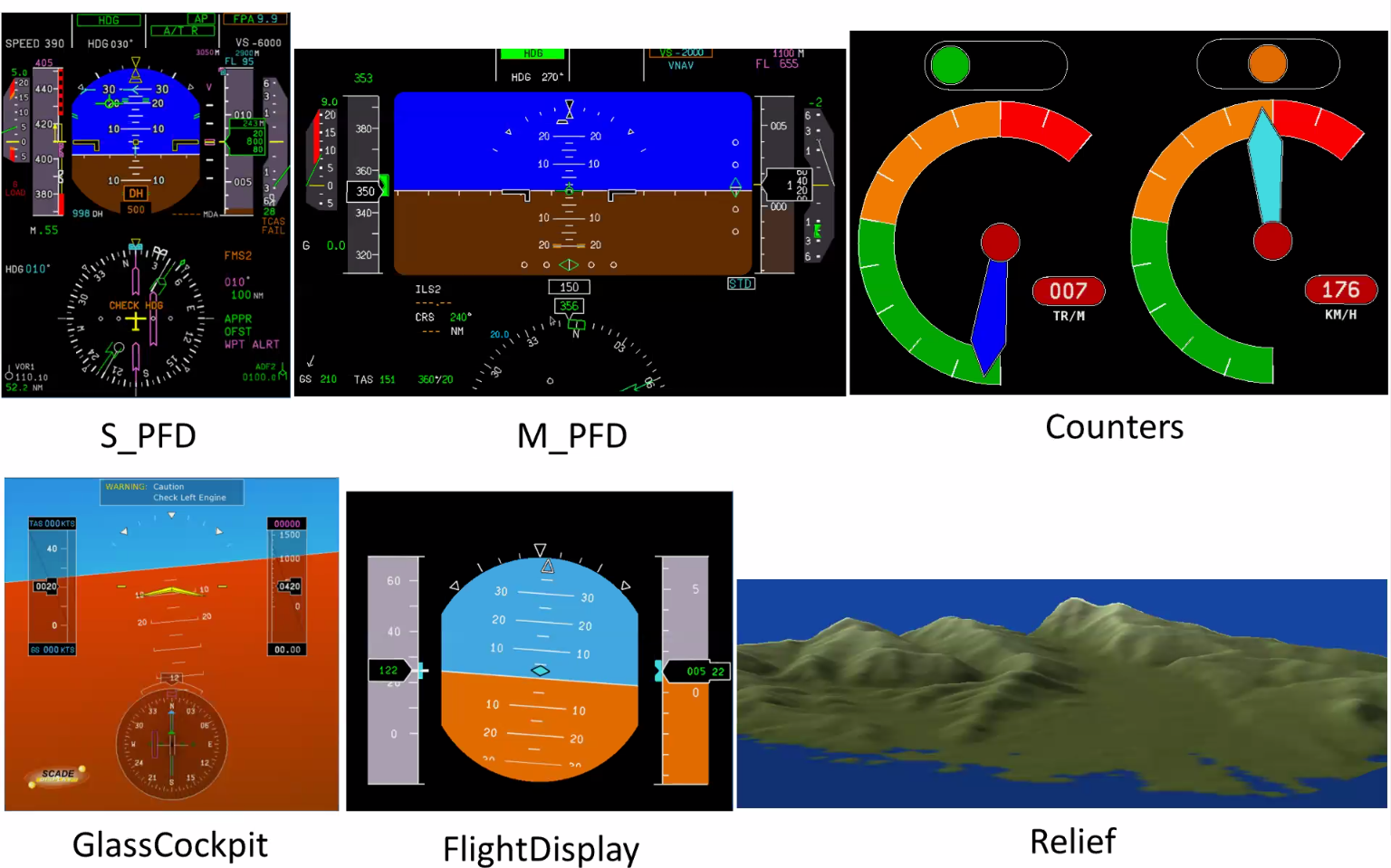}
    \caption{Examples used for testing by the KIAM.}
\end{figure}

It is possible to integrate the renderer with the ARINC~661 server for interactive applications as well as use directly to display complex non-interactive graphical applications that do not fit well with the available ARINC~661 widget system, such as cartography or terrain. The latter are becoming more common with the increasing capabilities of graphics cards and the trend to display more useful information with complex visual objects.

Applications use the OpenGL SC software rendering library. It is also possible to use the Vulkan SC software rendering library, which implementation remains for future work.

The results provided by the KIAM show that both implementations match the minimal requirements for performance and may be used on the pilot's display, but require either hardware or distributed rendering to display multiple or complex applications. The table below contains the minimal frame rate measurements performed by the KIAM on several test examples, shown on fig. 2, on one of the supported boards by JetOS. SWGL and HWGL in the columns refer to software and hardware OpenGL implementations, while 1 and 4 refer to the number of CPU cores used.

\begin{center}
    \begin{tabular}{ | m{0.12\textwidth} | m{0.08\textwidth} | m{0.08\textwidth} | m{0.08\textwidth} | }
        \hline
        & SWGL1 & SWGL4 & HWGL \\
        \hline
        S\_PFD & 5.9 & 13.8 & 10.8 \\
        \hline
        M\_PFD & 6.3 & 15.6 & 20.0 \\
        \hline
        Counters & 23.9 & 35.4 & 60.0 \\
        \hline
        GlassCockpit & 10.5 & 28.7 & 29.7 \\
        \hline
        FlightDisplay & 9.7 & 26.4 & 60.0 \\
        \hline
        Relief & 7.7 & 20.9 & 60.0 \\
        \hline
    \end{tabular}
\end{center}

The existing solution \cite{b9}, proposed by the KIAM, is tightly integrated with their applications and is difficult to make independent. It also lacks security mechanisms (for example, watchdogs). While in the current form we were unable to integrate it, this solution gave us an understanding of how our solution should look like.

This state of affairs leads to the fact that for efficient rendering of real-world applications we need to be able to build our graphical stack so that:
\begin{itemize}
    \item graphically independent widgets could be drawn in parallel and separately from each other;
    \item functionally independent graphical applications could have minimal influence on each other;
    \item graphical widgets and applications must simultaneously fit on the display.
\end{itemize}

By evaluating the performance of the currently available rendering code by the KIAM, we came to a conclusion that to support efficient multi-window management we need to bring another indirection level, namely, a compositor. By developing a software component that would provide a layer for creating graphical environments separately from the rendering stage we can not only solve the problems stated above, but additionally bring a number of improvements such as:

\begin{itemize}
    \item creating a unified API for software rendering between board support packages (BSP), well adapted for rendering ARINC~661 animations and widgets;
    \item ensuring sufficient software rendering performance when running multiple applications;
    \item ensuring that the rendering of one application does not interfere with the rendering of another for security reasons;
    \item additional tracking of problems that could arise at the rendering stage, and the ability to visually control them.
\end{itemize}

So, for instance, when adding any aircraft parameters to the indicator widget, such as speed, we can keep the same performance while providing parallelism at the hardware level.

% ARINC~653 standard does not provide dynamic memory management services, and as a result, implementations may not provide a way to free allocated memory, forcing all the memory to be assigned at the initialisation stage.

\section{JetOS graphics compositor}

\subsection{Unifying framebuffer access}

A framebuffer is a memory area for short-term storage of one or several frames in a digital form before sending them to a video output device. In general, a framebuffer has the following characteristics:
\begin{itemize}
    \item framebuffer width;
    \item framebuffer height;
    \item framebuffer pitch:
    \begin{itemize}
        \item the pitch of a framebuffer is simply the number of bytes that are in each row on the screen;
    \end{itemize}
    \item pixel format:
    \begin{itemize}
        \item R8G8B8A8;
        \item B8G8R8A8;
        \item A8R8G8B8;
        \item A8B8G8R8;
    \end{itemize}
    \item minimum and maximum frame rate;
    \item frame queue size (for buffered rendering).
\end{itemize}

To start solving the problem we defined an ARINC~653 system partition API for framebuffer access on the operating system level. This let us unify the approach applications interact with the target video adapter and also increased the efficiency of the rendering in some cases.

Our goal was to make the low-level API follow best industry practices, therefore we studied the implementations of similar pieces of software in non-safety-critical industries, such as Simple DirectMedia Layer (SDL) and the Unified Extensible Firmware Interface (UEFI) specification. In the first case, the offered interface is a lot more complex than required, and in the second case, it lacks essential functionality, for example, the presence of watchdogs, which are necessary to ensure safety, or image transformation. An important additional requirement was platform independence -- the core interface of the library should have no device-specific code.

\begin{figure}[H]
    \centering
\begin{verbatim}
typedef struct FRAMEBUFFER_CONTEXT_ {
    UINT32           width;
    UINT32           height;
    UINT32           pitch;
    FB_PIXEL_FORMAT  format;
    UINT32           framerate;
    UINT64           timeout;
    UINT32           queueDepth;
    VOID *           private;
} FRAMEBUFFER_CONTEXT;
\end{verbatim}
    \caption{Framebuffer context.}
\end{figure}

The API provides the ability to create a so-called framebuffer context, which describes the current state and parameters of the output surface. A simplified version of the framebuffer context is shown on fig. 3. Here \texttt{framerate} represents the minimal framerate, \texttt{format} is the currently used pixel format, and all other fields correspond to the list in the beginning of the section. All the fields are read-only and are modified inside the framebuffer implementation.

\begin{figure}[H]
    \centering
    \includegraphics[width=0.35\textwidth]{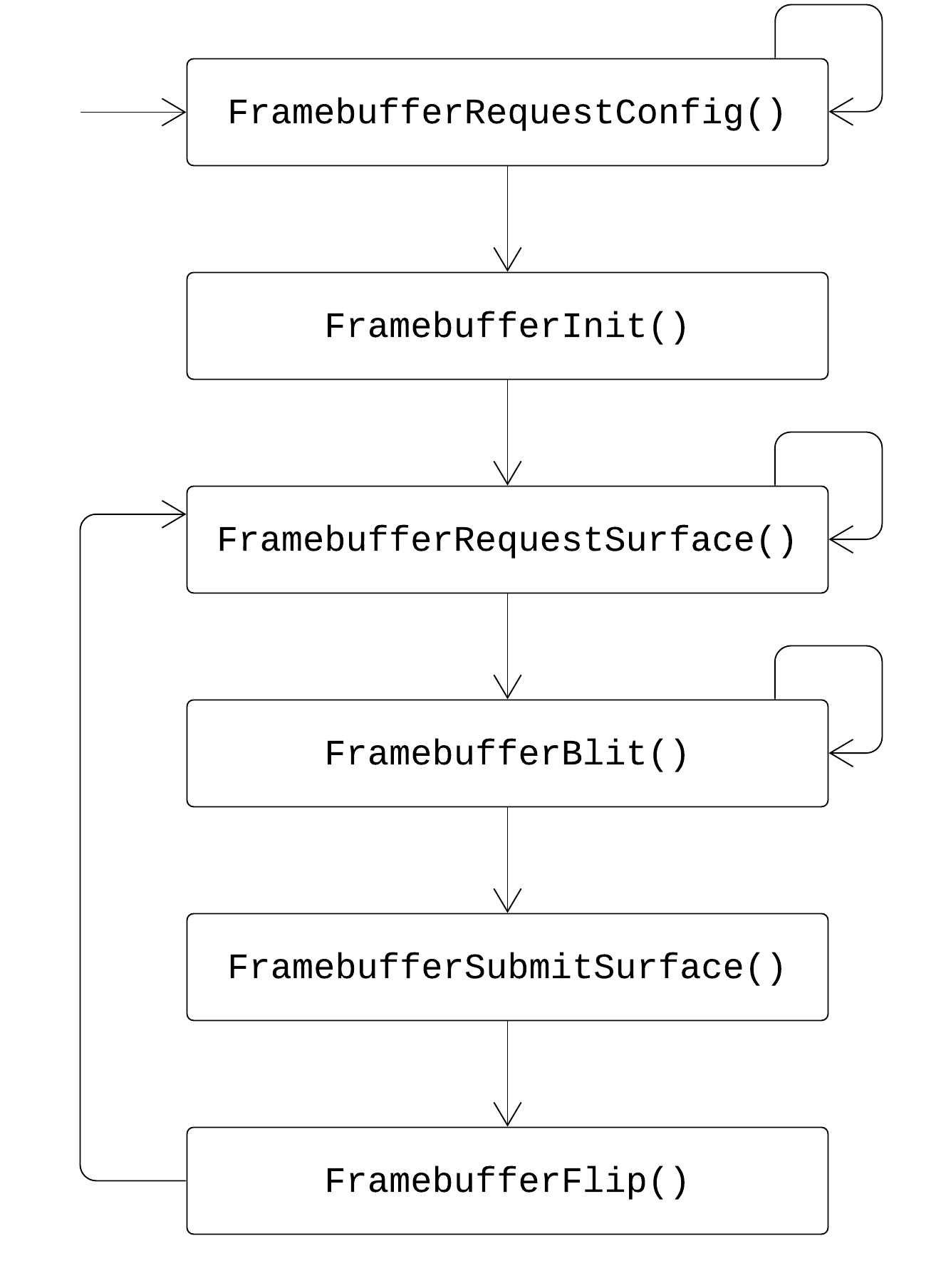}
    \caption{Framebuffer control flow.}
\end{figure}

Basic interaction happens through the methods exposed by every framebuffer implementation within the flow shown on fig. 5. The list includes:

\begin{itemize}
\item a method to request a free frame from the frame queue;
\item a method to fill the requested frame with data;
\item a method to send the filled frame to the frame queue;
\item a method to display the filled frame from the frame queue.
\end{itemize}

\subsection{Framebuffer compositor}

A compositor is a software component that composes windows on a display using application data from shared memory. It allows multiple application windows to exist on the display at the same time. It is also capable of performing various additional transformations (for example, converting pixels from one colour space to another). The compositor essentially is the only component that directly accesses GPU memory. The compositor generally has the following characteristics:

\begin{itemize}
    \item target screen width;
    \item target screen height;
    \item target screen pitch;
    \item minimum and maximum frame rate;
    \item frame data of the connected applications;
    \item pixel formats of the connected applications;
    \item target screen coordinates of the connected applications.
\end{itemize}

\begin{figure}[H]
    \centering
    \includegraphics[width=0.4\textwidth]{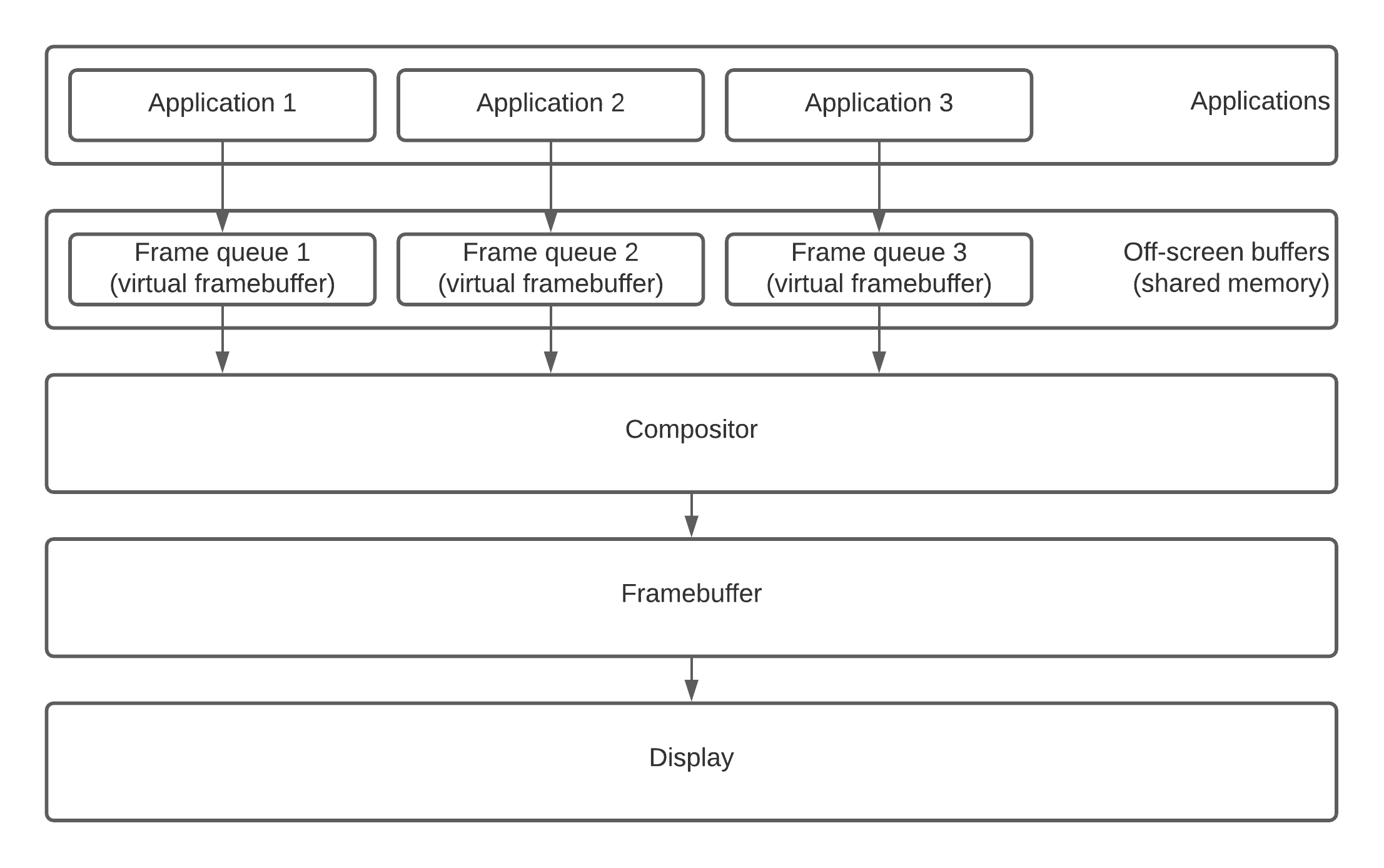}
    \caption{Visualisation of the data flow of three applications using compositor.}
\end{figure}

At the second step we provided a specialised virtual framebuffer implementation that would act as a normal framebuffer and be fully compatible with any framebuffer-built application, but would actually draw to a compositor. When composing into a framebuffer, each application sends its data to its own off-screen buffer. Several frames can be simultaneously in the buffer (there is a so-called frame queue). These frames are then passed to the compositor. It forms a single frame from them and renders it on the display using the framebuffer library as shown on fig. 3. The benefits of such approach are as follows:

\begin{itemize}
    \item when using compositor, applications are completely independent from each other (for example, to move a window to a different location, only the compositor code needs to be changed, or if there is a need to disable multiple applications, there is no need to change any code at all, only the configuration files);
    \item the application in which the error occurred does not affect other applications due to their independence:
    \begin{itemize}
        \item at the moment an error occurs in the application, the compositor disconnects it and continues to work, and the application itself has the ability to restart and continue to work;
    \end{itemize}
    \item the application's performance is also monitored:
    \begin{itemize}
        \item if the application's frame rate is below the minimum threshold, then the compositor disconnects it and continues to work, while the application itself turns off and has the ability to restart and continue to work;
    \end{itemize}
    \item there is a performance margin for the application:
    \begin{itemize}
        \item one can install a more complex application while maintaining overall performance (for example, if the application runs at 60 frames per second and the compositor runs at 30 frames per second, then the application could be replaced with a more complex application running at 30 frames per second without losing overall framerate).
    \end{itemize}
\end{itemize}

\begin{figure}[H]
    \centering
\begin{verbatim}
typedef struct FRAMEBUFFER_COMPOSITOR_ {
    _Atomic(BOOL)  ready;
    UINT32         magic;
    FB_PARAMS      parameters;
    UINT32         formatCount;
    UINT32         formatOffset;
    UINT32         frameCount;
    UINT32         frameOffset;
    UINT32         framePadding;
    UINT32         frameDataOffset;
    UINT32         privateOffset;
} FRAMEBUFFER_COMPOSITOR;
\end{verbatim}
    \caption{Framebuffer compositor client-server interface.}
\end{figure}

\begin{figure*}[t]
    \centering
    \includegraphics[width=0.9\textwidth]{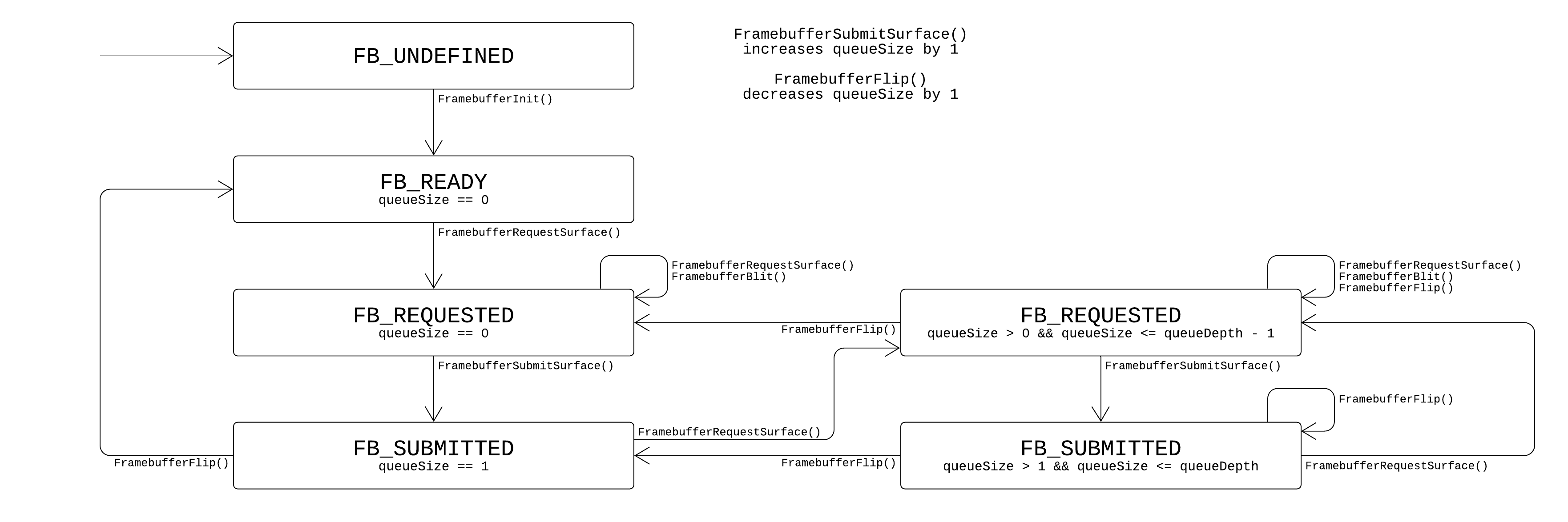}
    \caption{Framebuffer state machine with queuing.}
\end{figure*}

When developing a solution to this problem and adding new functionality, it was required not to degrade performance in comparison with a single framebuffer. The main problem here is effectively synchronising independently running applications and compositor.

Framebuffer compositor defines an interface between the server and each connected application over shared memory, which simplified version is shown on fig. 6. The client polls the \texttt{ready} field and then interprets the provided fields with pixel format, frame dimensions, and rate to build the list of supported configurations. Afterwards the offsets are used to address frame statuses (\texttt{frameOffset} field) and frame pixels (\texttt{frameData} field). To optimise memory copying each frame is aligned according to the hardware requirements exposed by the server through \texttt{framePadding} field.

\subsection{Frame buffering}

The common nature of all graphics pipelines, be that safety-critical software or video games, is that different frames can have different rendering costs. To allow to achieve higher frame rate in scenarios where some of the frames are known before they need to be shown to the viewer, we added frame queuing to our framebuffer API as shown on fig. 7. Each frame queue supports two different modes of operation:

\begin{itemize}
    \item display frames in turn order;
    \item clear the queue and display a new frame.
\end{itemize}

The separation is needed to make it possible to display critical frames, such as application disconnection, as early as possible, while maintaining tearing-free rendering in the other case. Each frame has a state that changes according to the state machine. Frames can be in the following states:

\begin{itemize}
    \item \texttt{FREE} -- the frame is free and can be retrieved by the application to fill it using the frame surface request function;
    \item \texttt{UPDATING} -- the frame is being prepared by the application for the server;
    \item \texttt{READY} -- the frame has been successfully prepared by the application for the server;
    \item \texttt{DRAWING} -- the frame is in use by the server.
\end{itemize}

The concept of a frame queue makes it possible to implement animations in a simpler way as frames for widgets are prepared and immediately placed in the queue without waiting for any commands. In fact, due to the presence of a compositor, which unavoidably results in off-by-one frame rendering, this is the only possible cost-effective solution in our case. Also, the queuing approach achieves double buffering and triple buffering mechanisms depending on queue size.

\subsection{Providing time guarantees}

To ensure that all frames arrive in time, watchdog timers are present both in framebuffer implementations and in the compositor implementation.

\begin{figure}[H]
    \centering
    \includegraphics[width=0.4\textwidth]{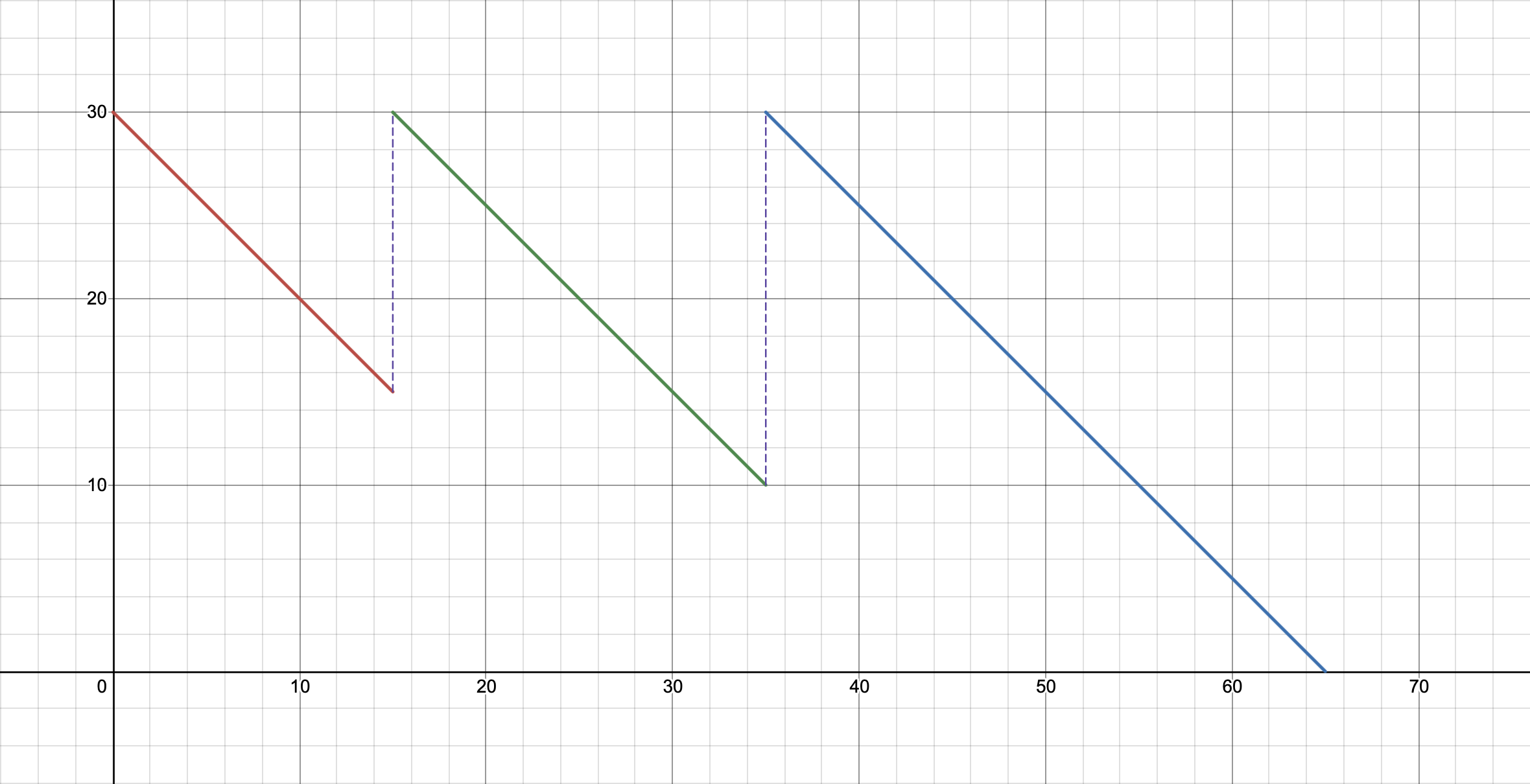}
    \caption{Watchdog timer operation scheme.}
\end{figure}

The application watchdog timer is a timer that is reset when the frame is fully rendered. With each frame rendered, the application's watchdog timer is reset to its initial value. When the timer expires, the Health Monitor is called. The scheme of the timer operation is shown on fig. 8. Different frames are marked with lines of different colours and the timer resets are marked with a dashed violet line. The vertical axis is the watchdog timer value and the horizontal axis is the time value.

With each frame the application renders, the server is updated with addresses for the application frames. If the application has not prepared its new frame, then no updates will occur and the old frame will not be freed. This allows to preserve the last rendered frame at no costs. On the other side, if the server does not receive a notification from the application within a certain time, then the application is forcibly disconnected from the server, which is made visible to the user. Its verification is performed by a separate process in the server partition.

%If the reset has not occurred within a certain time interval, the application is forced to disconnect from the server, visible to the user.

\subsection{Providing space isolation}

In accordance with ARINC~653 requirements, JetOS provides memory and time sharing between partitions on a constant cyclical basis. Each partition is allocated a certain fixed part of the total system operation period, thus ensuring the deterministic behaviour of the system. Each application and server runs on its own partition of the operating system. Images are transferred from the application to the server via shared memory blocks. Each application uses its own memory block to generate images. The server has read-only access to these blocks of memory.

The standard ARINC~653 interface supports multiple processor cores, but it is not used very often due to concerns that threads might interfere with each other. In addition, in the current version of the standard it is only possible to simultaneously run multiple processes within a single partition. While the restriction may eventually be lifted, currently two applications with separate address spaces cannot officially work in parallel as per ARINC~653 standard. To workaround this restriction in addition to the standardised symmetric multiprocessing (SMP) JetOS supports a so-called asymmetric multiprocessing (AMP) mode. That means, that on multicore systems JetOS supports the ability to run multiple modules (independent RTOS instances) on a single device. These modules work independently on different processor cores with minimal memory sharing and thus mostly disabled coherence, which leads to more predictable cache usage and better security and reliability, required by onboard software, compared to the SMP mode.

To improve the robustness of the project we chose the AMP mode when designing the framebuffer. On one side, it come at the cost of slightly lower efficiency than the SMP technology and limits the amount of fully independent clients by the amount of cores available on the board. Typical for aviation applications, the multicore PowerPC processors, like QorIQ\textsuperscript{\textregistered} P3041, have four cores. However, higher-end processors, like QorIQ\textsuperscript{\textregistered} P4080, with eight cores may well be used in aviation equipment for such specific tasks.

On another side, this architecture brings an additional security feature when designing an ARINC~661 server. The ARINC~661 standard neither protects from the point of resource consumption and bugs, nor does it protect against attacks on the rendering library. Using this interface, vendors could build a distributed ARINC~661 server, or any other widget abstraction layer, that may technically multiplex the pipeline again in a single drawing queue that will distribute drawing calls over the workers in separate address spaces with separate kernels. For legacy setups there also is a potential opportunity to create several ARINC~661 servers, so that each application works with its own, and these are additional advantages to reliability. Furthermore, in parallel with the ARINC~661 server, one can display complex graphic images on configurations without the ARINC~661 server, and still remain isolated.

\section{Achieved results}

The performance of one core was not enough to subsequently render more than one application at 30 frames per second or higher. To ensure that we solved this problem, we measured the performance of independently drawn SCADE widgets in compositor-enabled environment. For this we used the software rendering subsystem developed in the KIAM on real hardware with one of the samples, \texttt{counters}, from the previous chapter.

The results of testing \texttt{counters-framebuffer} (fig. 9) and \texttt{counters-compositor} (fig. 10) projects on QorIQ\textsuperscript{\textregistered} P3041 processor and Radeon\textsuperscript{\texttrademark} E4690 graphics adapter with compiler optimization -O3 are as follows (the applications run at 768×768 resolution, while the compositor runs at 1600×900 resolution):

\begin{itemize}
    \item \texttt{counters} -- 32 frames per second;
    \item \texttt{counters-framebuffer} -- 32.5 frames per second;
    \item \texttt{counters-compositor} -- 29.5 frames per second (the applications themselves run at 48 frames per second).
\end{itemize}

Here \texttt{counters} is the original project, \texttt{counters-framebuffer} is the same project after migration to the new framebuffer API, and \texttt{counters-compositor} is a project with two \texttt{counters-framebuffer} partitions running over the compositor. The results show that the new framebuffer API did not bring any performance downgrades, and the addition of the compositor adds a moderately negligible downgrade in overall performance, and leaves extra complexity room for each rendered application by boosting their performance by more than 15 frames per second.

\begin{figure}[H]
    \centering
    \includegraphics[width=0.4\textwidth]{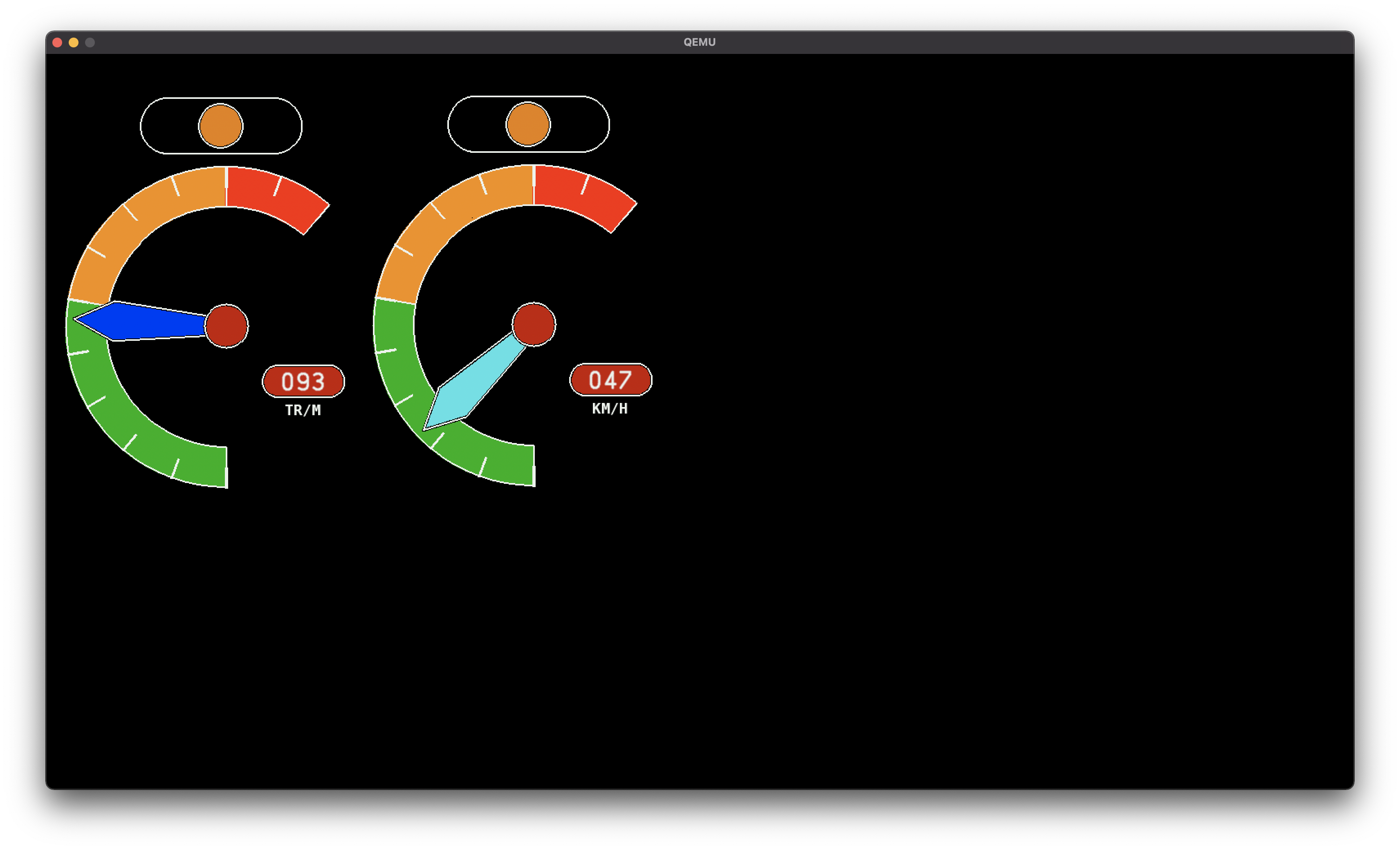}
    \caption{Testing \texttt{counters-framebuffer} project.}
\end{figure}

\begin{figure}[H]
    \centering
    \includegraphics[width=0.4\textwidth]{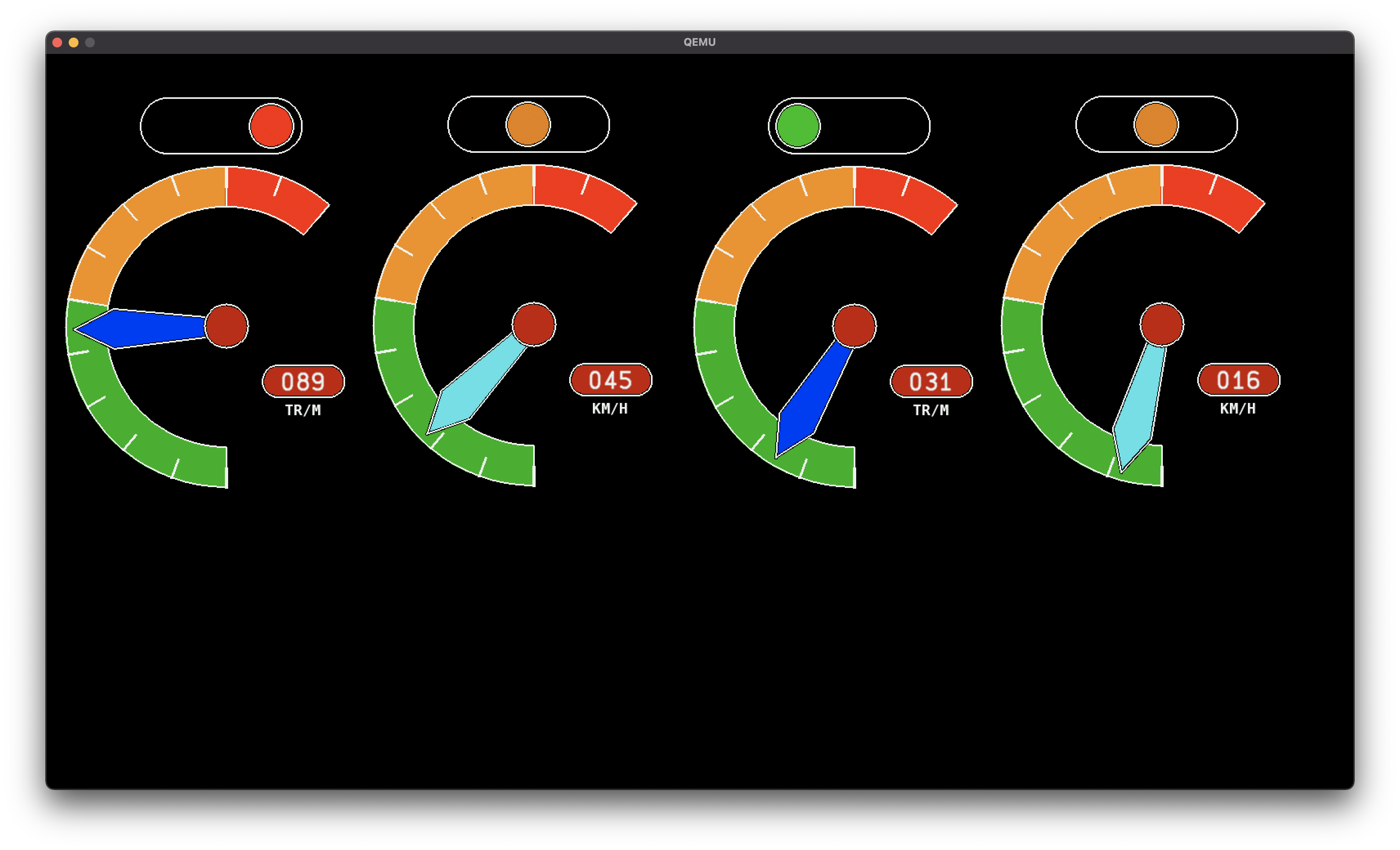}
    \caption{Testing \texttt{counters-compositor} project.}
\end{figure}

\section{Conclusion}

The development of a flight data visualisation system has its own specificity related to safety critical issues. This specificity often does not allow the use of ready-made, well-known solutions for one or another functionality. The existing solutions are mostly proprietary and there is very little information on them in the public domain. One of the reasons for this is that the implementations are very specific for each individual operating system. Also, quite a bit of information about the industry is published in academic journals. When developing software for onboard equipment, it is also necessary to take into account the fact that energy-saving and relatively low power processors are installed onboard. At the same time, the system must provide an interactive rendering speed.

\balance

The proposed solution to separate graphics pipeline to rendering and compositing stages on the operating system level has useful features that are generally absent in the existing solutions. Application software developers get unified rendering API, which follows best practices found in generic systems, like blitting or image transformations, and includes RTOS-specific measures like watchdogs to control the flow. It is possible to display multiple applications at once, reducing the rendering overhead of each and providing extra room to increase their complexity, with little to no overall rendering overhead. This is so as adding a compositor layer does not affect the performance of independently drawn objects if this can be parallelised across different cores. The use of asymmetric multiprocessing further pushes the edge of application independency by isolating not just the partitions but the entire operating system instances.

With this approach it becomes possible to create a secure ARINC~661 server where independent applications will not only be controlled but also rendered in independent partitions. Similarly the resource management of the graphic resources allocated for rendering of each application is enforced by providing separate quotas. This feature is not covered by the ARINC~661 standard, and so far we were unable to observe any similar solutions, despite this being in-line with the principles of ARINC~653 and all related airborne software equipment standards. Furthermore, it becomes additionally possible to combine ARINC~661, non-ARINC~661, or multiple ARINC~661 servers on one primary flight display (PFD) platform without the risk of any of them negatively affecting each other.

\section*{Acknowledgements}

We would like to thank our partners in the GosNIIAS and the KIAM, Boris Barladian and Nikolai Deriabin in particular, for their essential feedback and support during the research and development of this functionality.

\end{document}